\documentclass[10pt,twocolumn,letterpaper]{article}

\usepackage{cvpr}              

\usepackage{graphicx}
\usepackage{amsmath}
\usepackage{amssymb}
\usepackage{booktabs}
\usepackage{enumitem}
\usepackage[dvipsnames]{xcolor}
\usepackage{siunitx}
\usepackage[dvipsnames]{xcolor}
\usepackage[utf8]{inputenc}
\usepackage{tabularx} 

\usepackage[dvipsnames]{xcolor}

\definecolor{cvprblue}{rgb}{0.21,0.49,0.74}
\usepackage[pagebackref,breaklinks,colorlinks,citecolor=cvprblue]{hyperref}

\usepackage[capitalize]{cleveref}
\crefname{section}{Sec.}{Secs.}
\Crefname{section}{Section}{Sections}
\Crefname{table}{Table}{Tables}
\crefname{table}{Tab.}{Tabs.}

\def\paperID{*****} 
\def\confName{CVPR}
\def\confYear{2024}

\title{Hybrid Ensemble of Segmentation-Assisted Classification and GBDT for Skin Cancer Detection with Engineered Metadata and Synthetic Lesions from ISIC 2024 Non-Dermoscopic 3D-TBP Images}

\author{
Muhammad Zubair Hasan\\
University of North Texas\\
{\tt\small zubairhasan@my.unt.edu}
\and
Fahmida Yasmin Rifat\\
University of North Texas\\
{\tt\small fahmidayasminrifat@my.unt.edu}
}

\begin{document}
\maketitle

\begin{abstract}
Skin cancer is among the most prevalent and life-threatening diseases worldwide, with early detection being critical to patient outcomes. This work presents a hybrid machine and deep learning-based approach for classifying malignant and benign skin lesions using the SLICE-3D dataset from ISIC 2024, which comprises 401,059 cropped lesion images extracted from 3D Total Body Photography (TBP), emulating non-dermoscopic, smartphone-like conditions. Our method combines vision transformers (EVA02) and convolutional ViT hybrids (EdgeNeXtSAC) to extract robust features, employing a segmentation-assisted classification pipeline to enhance lesion localization. Predictions from these models are fused with a gradient-boosted decision tree (GBDT) ensemble enriched by engineered features and patient-specific relational metrics. To address class imbalance and improve generalization, we augment malignant cases with Stable Diffusion-generated synthetic lesions and apply a diagnosis-informed relabeling strategy to harmonize external datasets into a 3-class format. Using partial AUC (pAUC) above 80\% true positive rate (TPR) as the evaluation metric, our approach achieves a pAUC of 0.1755—the highest among all configurations. These results underscore the potential of hybrid, interpretable AI systems for skin cancer triage in telemedicine and resource-constrained settings.
\end{abstract}

\section{Introduction}
\label{sec:intro}
Skin cancer represents one of the most common forms of cancer worldwide, with its incidence continuing to rise, particularly in regions with high ultraviolet (UV) exposure. Early and accurate detection remains essential, as prognosis significantly improves when skin cancer is diagnosed in its initial stages~\cite{10.1007/978-3-031-72083-3_6}. However, traditional diagnostic procedures such as visual inspection by dermatologists and subsequent histopathological analysis are not only time-consuming but also subject to variability in clinical judgment, especially in cases involving subtle or atypical presentations~\cite{ding2023hi, SULEMAN20233533}. These limitations, coupled with growing patient loads, have fueled the development of automated and assistive diagnostic systems.

In recent years, deep learning (DL)–based computer-aided diagnosis (CAD) systems have demonstrated remarkable potential in dermatological imaging, offering faster and more reproducible assessments~\cite{baig2023light, Hamim2025SmartSkinXAI}. Among DL techniques, convolutional neural networks (CNNs) have emerged as the cornerstone of modern medical image analysis due to their ability to automatically extract hierarchical and discriminative features from raw lesion images~\cite{Mohadikar2023AdvancementIM, naqvi2023skin}. Several well-established architectures, including VGGNet~\cite{Lilhore2024}, ResNet~\cite{Khullar2025}, DenseNet~\cite{kousis2022deep}, and EfficientNet~\cite{tahir2023dscc}, have been successfully fine-tuned for binary and multi-class skin lesion classification tasks.

Given the limited size of annotated medical datasets, transfer learning has become a vital strategy, allowing researchers to adapt pre-trained models from large-scale datasets such as ImageNet to skin lesion data~\cite{srinivasu2021classification, ding2023hi}. Many studies have achieved near dermatologist-level accuracy using this paradigm~\cite{9014823}. More recently, ensemble and hybrid models have been employed to further enhance robustness and generalization by combining outputs from diverse CNNs or fusing them with traditional classifiers~\cite{Hamim2025SmartSkinXAI, behara2024improved}. Lightweight CNNs optimized for mobile and embedded deployment are also gaining attention~\cite{cheng2024enhanced, kousis2022deep}, expanding access to skin cancer screening in resource-constrained settings.

To address deeper feature learning and spatial modeling, attention mechanisms and capsule networks have been incorporated into CNN pipelines~\cite{behara2024improved, Lan_2023}. Furthermore, researchers are actively exploring domain adaptation~\cite{10.1007/978-3-031-72083-3_6} and unsupervised contrastive learning~\cite{9897435} to counter dataset shifts due to varying imaging protocols and skin tone diversity. Despite these advancements, several critical challenges persist: significant class imbalance, high intra-class similarity between lesion types, and the lack of generalizability across heterogeneous populations~\cite{Khullar2025}. The development of explainable AI (XAI) techniques also remains vital to enhance clinician trust and ensure responsible clinical integration.

Our contributions in this study are as follows:

\begin{itemize}
    \item We introduce a \textbf{segmentation-assisted classification strategy} to enhance lesion boundary detection and improve overall accuracy in cancer detection for the image based models. This is the main contribution of our work.
    
    \item We propose a \textbf{hybrid ensemble framework} combining Gradient Boosting Decision Trees (GBDT) and image-based deep learning models (e.g., EVA02, EdgeNeXt), allowing complementary exploitation of tabular and visual features for robust skin cancer classification. EVA02, EdgeNeXt were chosen based on the performance per parameter count factor.

    \item We incorporate \textbf{patient-specific relational features} by analyzing relative lesion characteristics (e.g., outlier scores, lesion-to-patient ratios), enriching the feature space with contextual insights for better inter-lesion discrimination.

    \item We utilize \textbf{external and synthetic data augmentation}, using stable diffusion generated malignant lesions to mitigate class imbalance and improve generalization under scarce positive examples.

    \item We introduce a \textbf{diagnosis-informed data integration strategy} by relabeling external datasets into a simplified 3-class setup (nevus, melanoma, bkl) using standardized diagnosis mappings, thereby enabling pretrained models to provide transferable representations for downstream lesion classification.
\end{itemize}

    



\section{Methodology}
\subsection{Dataset}

For this project, we utilize the dataset provided in the \textbf{ISIC 2024 - Skin Cancer Detection with 3D-TBP} competition, hosted by the International Skin Imaging Collaboration (ISIC) on the Kaggle platform\cite{Kurtansky2024}. The dataset consists of diagnostically labeled skin lesion images extracted from 3D Total Body Photography (TBP) systems. These images emulate non-dermoscopic, smartphone-like photography, aligning them with teledermatology use cases. 

The dataset comprises a total of \textbf{401,059 cropped lesion images}, each approximately \textbf{128$\times$128 pixels} in size. These are stored in JPEG format and are accompanied by rich metadata stored in CSV files. The metadata captures diverse types of information categorized as follows: \textbf{Patient Demographics}, \textbf{Lesion Diagnosis Information}, \textbf{Lesion Location}, \textbf{Lesion Size \& Geometry}, \textbf{Lesion Shape \& Symmetry}, \textbf{Lesion Location Coordinates}, and \textbf{Lesion Color Information}. The dataset is structured into several key components:
\begin{itemize}
    \item \textbf{SLICE-3D Dataset}: The complete collection of cropped lesion images from 3D TBP, including diagnostic malignancy annotations.
    \item \textbf{Train Images and Metadata}: A labeled subset of images with corresponding metadata used for supervised training.
    \item \textbf{Test Images and Metadata}: An unlabeled set of images with metadata, used for model inference and evaluation.
\end{itemize}

To address the severe class imbalance inherent in the original dataset, we augment the training set with synthetic malignant examples generated using \textbf{Stable Diffusion}, a generative image synthesis technique. The final class distribution is as follows: \textbf{Malignant – 393 real lesions + 30,228 synthetic lesions}, and \textbf{Benign – 400,666 lesions}. This synthetic augmentation significantly aids in improving model robustness and generalization. To illustrate the dataset, we present a sample visualization of the skin lesion images used for training. The following figure shows an example from the dataset:

\begin{figure}[h]
    \centering
    \includegraphics[width=0.9\linewidth]{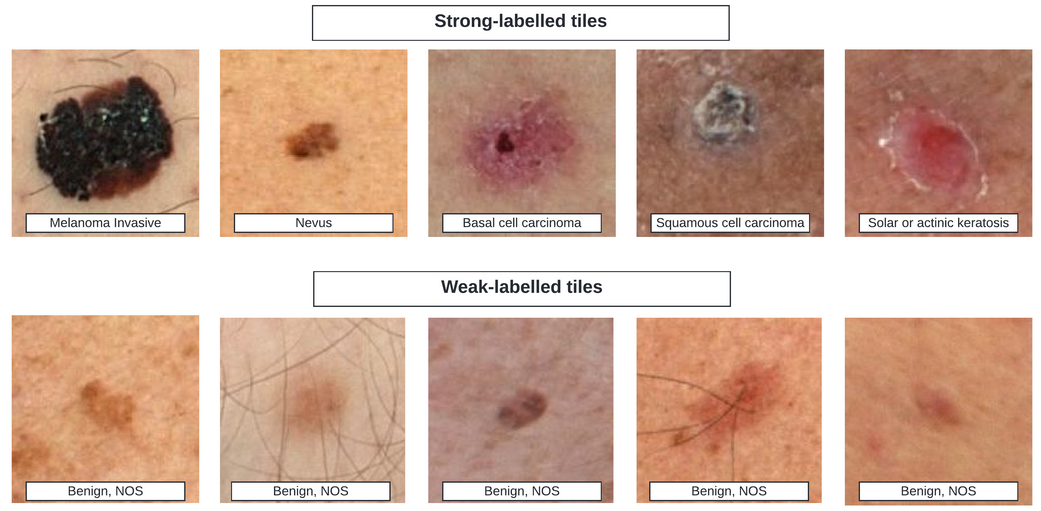}
    \caption{Sample images from the SLICE-3D dataset. These cropped lesion images are extracted from 3D TBP, mimicking non-dermoscopic conditions.}
    \label{fig:sample_data}
\end{figure}
\subsection{Our Approach}
\label{sec:method}
\begin{figure*}[ht]
  \centering
  \includegraphics[width=0.8\textwidth]{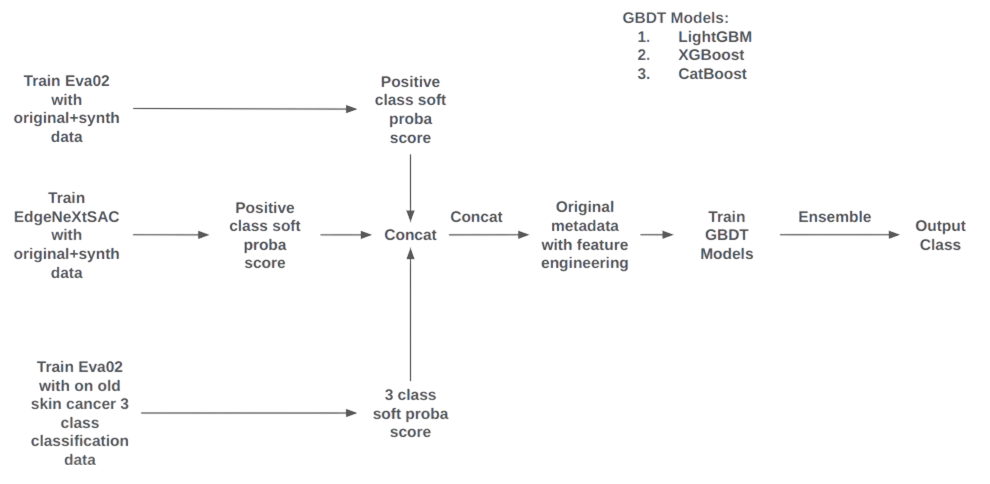} 
  \caption{Overall pipeline: synthetic malignant lesions are added to real data. Multiple image models (EVA02, EdgeNeXtSAC, EVA02-old) are trained and ensembled. Their probability outputs are concatenated with metadata and engineered features to train a robust GBDT ensemble.}
  \label{fig:pipeline}
\end{figure*}
Our approach to skin lesion classification integrates multiple levels of information — raw images, handcrafted features, and external model predictions — to enhance robustness and generalization, especially in imbalanced and real-world scenarios. The full pipeline is summarized in Figure~\ref{fig:pipeline}.





\begin{table}[!t]
  \centering
  \caption{Feature groups used by GBDT models.}
  \label{tab:feat_types}
  \begin{tabular}{p{0.35\linewidth} p{0.1\linewidth} p{0.45\linewidth}}
    \toprule
    \textbf{Feature Group} & \textbf{Count} & \textbf{Example Feature} \\
    \midrule
    Raw numeric & 34 & \texttt{clin\_size\_long\_diam\_mm}, \texttt{tbp\_lv\_H} \\
    Raw categorical (1-hot) & 6 & \texttt{sex}, \texttt{anatom\_site\_general} \\
    One-hot Categorical Features & 47 \\
    Engineered features & 43 & \texttt{lesion\_shape\_index}, \texttt{border\_complexity} \\
    Patient-normalized features & 76 & \texttt{tbp\_lv\_H\_patient\_norm}, \texttt{lesion\_size\_ratio\_} \newline \texttt{patient\_norm} \\
    Patient-aggregated metrics & 3 & \texttt{count\_per\_patient}, \texttt{tbp\_lv\_areaMM2\_bp} \\
    External model predictions & 5 & \texttt{predictions\_eva}, \texttt{predictions\_edg} \\
    \midrule
    \textbf{Total} & \textbf{214} & -- \\
    \bottomrule
  \end{tabular}
\end{table}

\begin{figure*}[ht]
  \centering
  \includegraphics[width=0.8\textwidth]{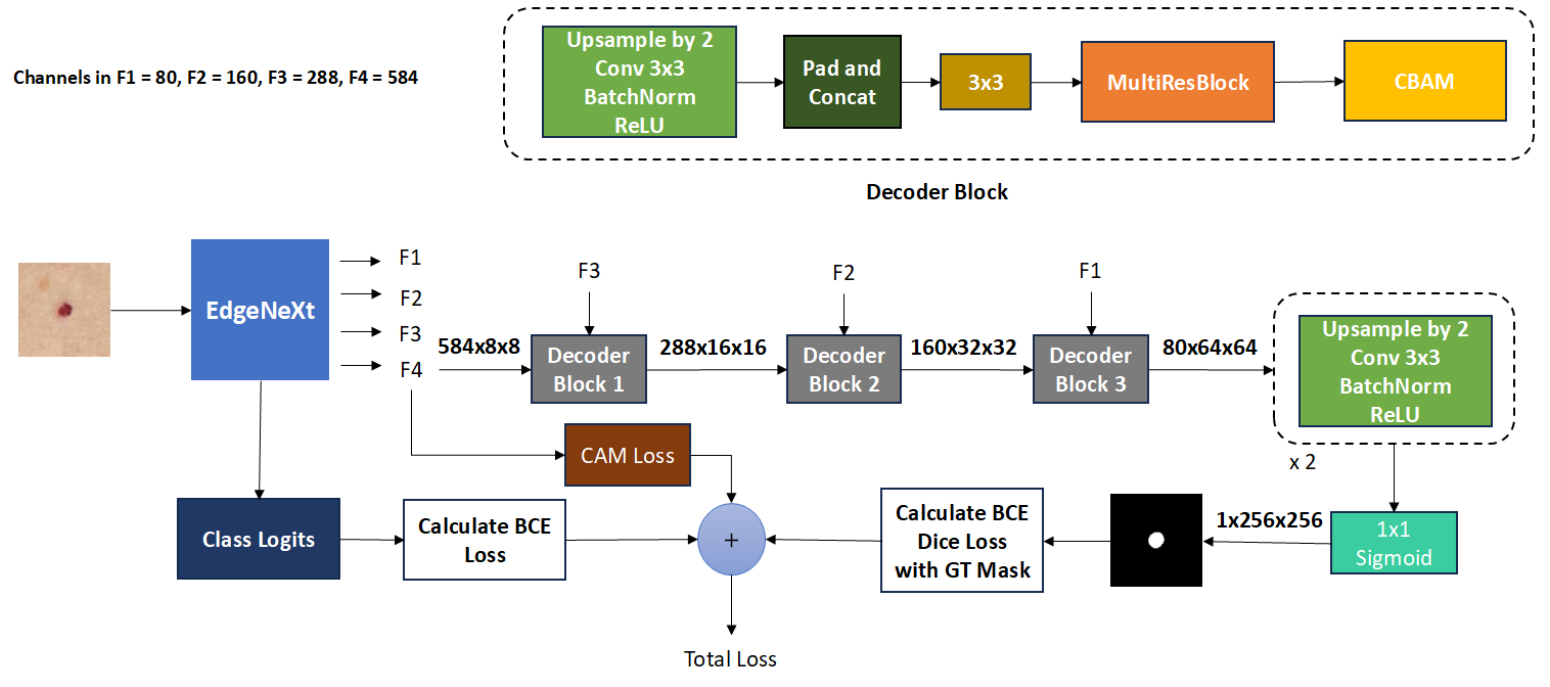} 
  \caption{Segmentation-assisted classification architecture (EdgeNeXtSAC). The backbone (EdgeNeXt) feeds multi-scale features into a lightweight decoder block with attention (CBAM). Both BCE and Dice losses guide lesion-focused segmentation and classification jointly.}
  \label{fig:edgenext_sac}
\end{figure*}

\subsubsection{Image-Based Classification}
\label{sec:image_models}

We train two convolutional architectures to generate soft predictions: EVA02\cite{Fang_2024} and our EdgeNeXt\cite{maaz2022edgenext} inspired EdgeNeXtSAC.

\paragraph{EVA02.}

We employ the EVA02-small vision transformer architecture as a high-capacity image encoder for skin lesion classification. This model, pre-trained on ImageNet-22k and fine-tuned on ImageNet-1k, was further adapted to our binary classification task distinguishing between benign and malignant lesions. To ensure robust evaluation and prevent data leakage across patients, we use a \textbf{3-fold Stratified Group K-Fold cross-validation} strategy. Each fold preserves the malignancy class ratio while enforcing patient-level disjoint partitions between training and validation sets. This setup guarantees that no images from the same patient appear in both the training and validation folds, which is essential in medical imaging where multiple lesions may exhibit correlated appearance due to shared patient-level characteristics.

To mitigate the severe class imbalance in the original dataset, training batches were formed using a 1:1 sampling ratio between benign and malignant samples. This oversampling strategy ensures that each training batch includes a balanced number of positive and negative samples, enabling more stable convergence and improved sensitivity to malignant lesions. The model was trained using the \textbf{binary cross-entropy loss} function, with early stopping based on validation performance to avoid overfitting.

In addition to the binary classifier, we trained a second EVA02 model as a \textbf{3-class classifier} (melanoma / nevus / keratinocyte lesions) using external data curated from previous ISIC competitions. To ensure consistency and reduce label noise, we re-mapped the original diagnostic labels into three clinically meaningful categories. Lesions labeled as \texttt{melanoma} were assigned to the melanoma class, while those labeled as \texttt{nevus} were retained in the nevus class. Keratinocyte-related lesions—including \texttt{basal cell carcinoma}, \texttt{seborrheic keratosis}, \texttt{solar lentigo}, and \texttt{lentigo NOS}—were grouped together under a unified \texttt{bkl} (benign keratinocyte lesion) class. All remaining benign lesions with diagnoses not included in the above keratinocyte mapping were also reclassified under the nevus category. This label harmonization allowed us to construct a well-structured 3-class training dataset that captures the key lesion types while smoothing diagnostic variance. The resulting 3-class classifier provided auxiliary outputs that served as input features for downstream GBDT models, contributing valuable intermediate representations to help distinguish borderline or ambiguous cases.

This relabeling strategy consolidates the multi-label diagnostic space into semantically meaningful categories aligned with clinical expectations. It also provides an intermediate diagnostic signal for borderline lesions that are otherwise underrepresented in binary training. Both versions of EVA02—the binary and the 3-class classifier—were trained using the same 3-fold Stratified Group K-Fold setup to ensure consistent patient separation and reproducibility.

The soft class probabilities output by both models were extracted for each lesion and subsequently included as input features to the GBDT ensemble. This design allows the downstream tabular model to incorporate high-level visual cues learned from powerful vision transformers, contributing to more informed and discriminative lesion classification.

\paragraph{EdgeNeXtSAC.}

To enhance lesion localization and improve classification performance, we designed a \textbf{Segmentation-Assisted Classifier (EdgeNeXtSAC)} by extending the EdgeNeXt architecture with a dual-head framework. The model jointly performs lesion classification and segmentation by leveraging shared encoder features and task-specific decoders. The overall architecture is illustrated in Figure~\ref{fig:edgenext_sac}.

\textbf{Encoder Backbone.} We use EdgeNeXt as the encoder, which extracts hierarchical feature maps at four stages (\texttt{F1} to \texttt{F4}). These features encode increasingly abstract representations of the input skin lesion and serve as skip connections for a UNet-inspired decoder.

\textbf{Segmentation Decoder.} The decoder progressively upsamples the deep feature maps using a novel multi-stage design. Each \textbf{Decoder Block} (see Figure~\ref{fig:edgenext_sac}) begins with an upsampling module that doubles the spatial resolution of the input using a sequence of operations: a $2\times$ upsample followed by a $3\times3$ convolution, batch normalization, and ReLU activation. This is followed by a skip-connection fusion stage, where the upsampled feature map is padded and concatenated with the corresponding encoder feature of the same scale to preserve spatial context. The concatenated feature is then passed through a $3\times3$ convolution layer to integrate information from both sources. Next, a \textbf{MultiResBlock} \cite{IBTEHAZ202074} captures multi-scale lesion features through a set of parallel convolutions of varying kernel sizes, enabling the model to learn both fine- and coarse-grained patterns. Finally, a \textbf{Convolutional Block Attention Module (CBAM)} \cite{woo2018cbam} is applied to adaptively refine feature representations using channel-wise and spatial attention, promoting the learning of salient lesion boundaries while suppressing irrelevant background signals. The final segmentation output is passed through a $1\times1$ convolution followed by a \texttt{sigmoid} activation to produce a lesion probability map of size $1\times256\times256$.

\textbf{Loss Functions.} The total loss $\mathcal{L}_{\text{total}}$ is a weighted sum of three components:

1. \textbf{Classification Loss (BCE):} Calculated between predicted class logits and binary ground-truth labels using Binary Cross Entropy:
\begin{equation}
\mathcal{L}_{\text{cls}} = -[y \cdot \log(\hat{y}) + (1 - y) \cdot \log(1 - \hat{y})]
\end{equation}
where $y \in \{0, 1\}$ is the true class label and $\hat{y}$ is the predicted probability of the lesion being malignant.

2. \textbf{CAM Loss (Dice):} A weakly-supervised segmentation loss that aligns the Class Activation Map (CAM) from the classification head with the ground-truth segmentation mask:
\begin{equation}
\mathcal{L}_{\text{cam}} = 1 - \frac{2 \sum M_{\text{cam}} \cdot M_{\text{gt}}}{\sum M_{\text{cam}} + \sum M_{\text{gt}} + \epsilon}
\end{equation}
where $M_{\text{cam}}$ is the normalized CAM and $M_{\text{gt}}$ is the binary ground-truth mask. This encourages the classification network to activate over relevant lesion regions.

3. \textbf{Segmentation Loss (BCE + Dice):} The final segmentation output is supervised using a compound loss combining pixel-wise Binary Cross Entropy and Dice Loss:
\begin{equation}
\mathcal{L}_{\text{seg}} = \mathcal{L}_{\text{bce}}^{\text{mask}} + \mathcal{L}_{\text{dice}}^{\text{mask}}
\end{equation}

4. \textbf{Total Loss:} The three losses are summed to form the overall training objective:
\begin{equation}
\mathcal{L}_{\text{total}} = \mathcal{L}_{\text{cls}} + \mathcal{L}_{\text{cam}} + \mathcal{L}_{\text{seg}}
\end{equation}

This multi-task setup enforces consistency between classification and segmentation, guiding the model to focus on spatially meaningful regions. It also improves robustness, especially in scenarios where lesion boundaries are subtle or labels are weak. The predicted class logits from EdgeNeXt and the intermediate attention maps are subsequently used as features in the final GBDT ensemble.

\subsubsection{Feature Engineering}
To enhance the representational power of metadata and improve classification performance, we extended the original metadata with a rich set of \textbf{174 engineered features}, resulting in a total of \textbf{214 input features} for our gradient-boosted decision tree (GBDT) models (see Table~\ref{tab:feat_types}). Our feature engineering strategy was designed to capture both lesion-level information and broader patient-level context, while also integrating insights from deep learning models trained on image data.

First, we introduced geometric and color-based descriptors that capture lesion shape, asymmetry, and color contrast. These include features such as the \texttt{lesion\_shape\_index}, \texttt{hue\_contrast}, and \texttt{lesion\_color\_difference}, all of which are informative for distinguishing malignant from benign lesions. Next, to mitigate inter-subject variability and standardize lesion comparisons, we computed \textbf{patient-normalized features} by aggregating values such as lesion size or hue across each patient and then calculating relative deviations for individual lesions. This approach helps the model identify lesions that are atypical for a specific patient, thereby improving generalization. We also incorporated \textbf{aggregated lesion metrics} by anatomical site and patient groupings to enrich spatial context. Categorical variables such as \texttt{sex} and \texttt{anatomical\_site} were one-hot encoded, resulting in a significant number of binary indicator features that preserve non-ordinal relationships.

In addition to engineered features derived from metadata, we embedded \textbf{predictions from image-based deep learning models} (EVA02 and EdgeNeXt-SAC) as input features to the GBDT models. This fusion of structured and unstructured data allows GBDT models to benefit from learned visual representations. Moreover, we used a \textbf{3-class classifier (nevus/melanoma/bkl)} trained on external datasets with standardized diagnosis mappings to enrich the GBDT inputs. These outputs, mapped back to binary malignancy labels, provided an additional perspective on lesion characterization, improving model sensitivity to borderline cases. To prevent overfitting of GBDT models to highly confident deep model predictions, we followed a regularization-inspired approach by injecting Gaussian noise (\(\sigma = 0.1\)) into the external prediction features. This simple yet effective technique reduces model over-reliance on any single prediction channel and encourages learning more robust decision boundaries.

\subsubsection{Gradient Boosted Decision Trees (GBDTs)}
\label{sec:gbdt}

While convolutional neural networks (CNNs) effectively capture spatial cues from lesion images, we complement them with a suite of gradient-boosted decision tree (GBDT) models—LightGBM, XGBoost, and CatBoost—trained on a rich set of engineered features to enhance interpretability and probabilistic calibration. To prevent data leakage and maintain patient independence across folds, we employ stratified GroupKFold cross-validation with three different random seeds, resulting in a 3×5-fold setup. Within each fold, we address class imbalance by oversampling rare malignant cases (e.g., melanoma) and undersampling the abundant benign cases, ensuring a balanced distribution during training. The final classification is derived through a two-stage ensemble framework: first, we concatenate the softmax outputs from image-based models (binary EVA02, multi-class EVA02, and EdgeNeXtSAC) with the handcrafted tabular features; then, an ensemble of 45 GBDT models (spanning 3 model types, 5 folds, and 3 seeds) aggregates their predictions to generate the final malignancy score. This design synergistically integrates deep visual representations, synthetic data augmentation, and structured metadata for robust and generalizable skin cancer classification.

\section{Evaluation Metric}

The performance of the proposed model is evaluated using the \textbf{partial Area Under the ROC Curve (pAUC)} above an 80\% true positive rate (TPR). This metric is specifically designed to focus on clinically relevant regions of the ROC curve where high sensitivity is crucial.

\subsection{Receiver Operating Characteristic (ROC) Curve}
The ROC curve illustrates the trade-off between the true positive rate (TPR) and the false positive rate (FPR) at various classification thresholds. A classifier's performance is typically summarized using the area under the ROC curve (AUC), where a higher value indicates better discrimination between malignant and benign lesions.

\subsection{Partial AUC (pAUC) Above 80\% TPR}
Unlike traditional AUC, which considers the entire ROC curve, the ISIC 2024 competition focuses on the partial AUC (pAUC) in the region where TPR is at least 80\%. This ensures that models prioritize high sensitivity, which is critical for medical applications where missing a malignant lesion can have severe consequences. The pAUC is computed as follows:

\begin{equation}
\text{pAUC} = \int_{0.8}^{1.0} \text{ROC}(t) \, dt
\end{equation}

where \( \text{ROC}(t) \) represents the ROC curve function and the integral is calculated over the FPR range corresponding to TPR values greater than or equal to 80\%.

\subsection{Scoring Range}
The pAUC score is normalized to a range of [0.0, 0.2], where higher values indicate better model performance in the clinically relevant high-sensitivity region.

\subsection{Clinical Relevance}
This evaluation metric aligns with real-world diagnostic needs, where a high true positive rate is essential for early cancer detection. By emphasizing performance in the high-sensitivity region, this metric encourages the development of models suitable for triaging applications in telemedicine and non-specialized settings.

\section{Experimental Results and Findings}
\label{sec:exp}

We evaluate our proposed pipeline under multiple configurations and dissect the results using partial AUC (pAUC), model ensembles, interpretability tools, and segmentation-aided visualization. 

\subsection{Model Performance and Ensemble Gains}

Table~\ref{tab:image_model_scores} presents the partial AUC (pAUC) scores of individual image-based classifiers and their ensembles, with and without synthetic augmentation. The EVA02 transformer-based model, when trained on original data alone, yields a pAUC of 0.1516. However, incorporating synthetically generated malignant lesions significantly boosts performance to 0.1633. This highlights the utility of data augmentation—particularly in addressing the extreme class imbalance and enhancing model sensitivity to rare malignant classes.

For the CNN-based EdgeNeXt architecture, the baseline classification model without segmentation guidance achieves a pAUC of 0.1401. Introducing segmentation supervision through our proposed EdgeNeXtSAC variant increases this to 0.1439. Further improvement is observed when synthetic malignant samples are included, reaching 0.1576. This confirms that segmentation-assisted learning helps the model localize critical lesion regions more effectively, and that synthetic data expands the decision boundary to better generalize on borderline or underrepresented malignant cases.

\begin{table}[ht]
  \centering
  \caption{Performance (pAUC) of image models with and without synthetic data.}
  \label{tab:image_model_scores}
  \begin{tabular}{l c}
    \toprule
    \textbf{Model} & \textbf{pAUC} \\
    \midrule
    EVA02 (real only) & 0.1516 \\
    EVA02 + Synth & 0.1633 \\
    EdgeNeXt (real only) & 0.1401 \\
    EdgeNeXtSAC (real only) & 0.1439 \\
    EdgeNeXtSAC + Synth & 0.1576 \\
    \bottomrule
  \end{tabular}
\end{table}

Table~\ref{tab:ensemble_results} extends this analysis to feature-based gradient boosting models. A baseline ensemble of GBDT models trained solely on raw metadata features scores 0.1500. Applying our feature engineering pipeline—including patient-normalized features, lesion-level outlier metrics, and spatially contextual aggregations—improves this to 0.1644, surpassing the image-only model performances. Finally, integrating softmax probability scores from the trained image classifiers as additional input features results in a peak performance of 0.1755. This highest score confirms the synergy between structured metadata, feature-rich handcrafted attributes, and high-level visual embeddings learned by deep networks.

\begin{table}[ht]
  \centering
  \caption{GBDT ensemble performance with and without feature engineering and image model fusion.}
  \label{tab:ensemble_results}
  \begin{tabular}{l c}
    \toprule
    \textbf{Configuration} & \textbf{pAUC} \\
    \midrule
    GBDT Ensemble (raw) & 0.1500 \\
    + Feature Engineering & 0.1644 \\
    + Image Model Probabilities & \textbf{0.1755} \\
    \bottomrule
  \end{tabular}
\end{table}

\subsection{Prediction Confidence Analysis}
\begin{figure*}[ht]
  \centering
  \includegraphics[width=\textwidth]{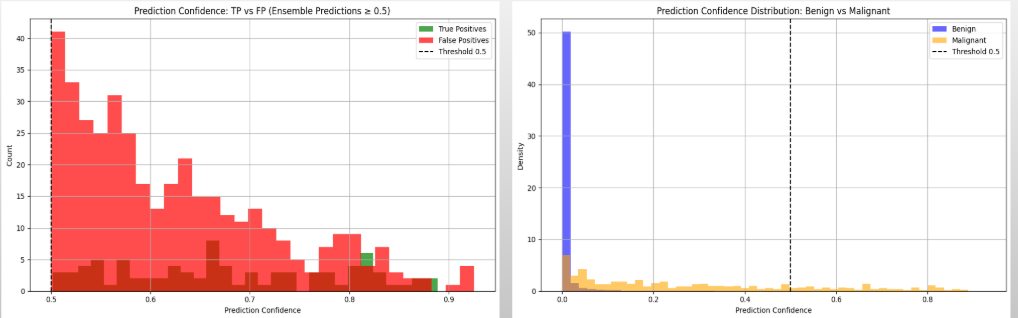}
  \caption{Left: Distribution of true positives (green) vs. false positives (red) across prediction confidence. Right: Overall predicted confidence for benign vs. malignant classes. Benign samples are confidently predicted near 0; malignant cases show greater uncertainty.}
  \label{fig:conf_analysis}
\end{figure*}
Figure~\ref{fig:conf_analysis} provides a two-part analysis of the ensemble model’s prediction confidence for benign versus malignant lesions. In the left panel, we examine how true positives (TPs) and false positives (FPs) are distributed across the prediction confidence range, conditioned on predicted scores $\geq 0.5$. We observe that true positives—correctly identified malignant lesions—tend to occur predominantly in the 0.6 to 0.9 confidence range. This indicates that when the model is confident in predicting malignancy, it is often correct. In contrast, false positives—benign lesions misclassified as malignant—are densely concentrated near the decision threshold at 0.5. This pattern reveals that most classification errors occur in a region of low confidence, suggesting these are “soft errors” that might be mitigated through threshold adjustment or calibration techniques such as Platt scaling or isotonic regression. The sparse number of FPs at higher confidences further reinforces that the model rarely misclassifies benign lesions with strong certainty.

The right panel illustrates the full prediction confidence distributions for both benign and malignant samples. Benign predictions (blue) form a sharp spike near zero, showing that the model has high certainty when predicting the negative class. This heavy left skew indicates excellent model calibration for benign cases. Malignant predictions (orange), however, are spread more uniformly across the full confidence range from 0 to 1, reflecting the ambiguity and difficulty of detecting malignant lesions. Many malignant samples are predicted with confidence below the decision threshold of 0.5, explaining some of the false negatives. Only a subset of malignant cases are classified with high confidence, aligning with the true positive concentration observed in the left plot.

\subsection{Segmentation-aided Classification Visualizations}
\begin{figure}[ht]
  \centering
  \includegraphics[width=\linewidth]{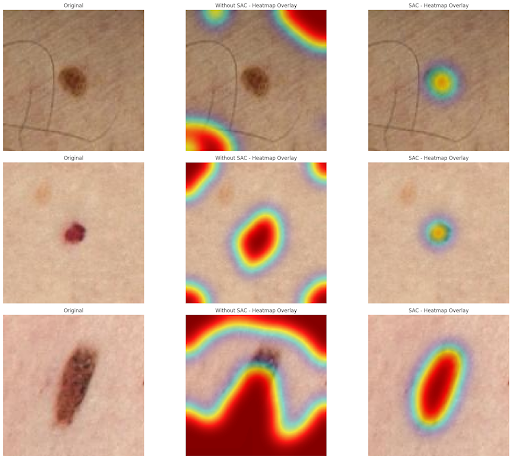}
  \caption{Comparison of lesion localization using GradCAM++ overlays. Each row shows: (1) the input image, (2) GradCAM++ heatmap from the baseline EdgeNeXt model (without SAC), and (3) GradCAM++ heatmap from our proposed EdgeNeXtSAC model. The SAC-enhanced model consistently focuses on the lesion regions with higher spatial precision, highlighting its improved localization and cancer detection capabilities.}
  \label{fig:seg_steps}
\end{figure}

Figure~\ref{fig:seg_steps} presents a visual comparison between the baseline EdgeNeXt model and our proposed EdgeNeXtSAC model using GradCAM++ attention overlays. Each row displays the original input image, followed by the GradCAM++ attention heatmap from the vanilla EdgeNeXt model and the GradCAM++ heatmap generated by EdgeNeXtSAC. While the baseline model often highlights diffuse or irrelevant regions, our SAC-enhanced model exhibits sharply localized attention precisely over the lesion area. This demonstrates the effectiveness of segmentation-guided learning in steering the model's focus toward clinically meaningful regions, contributing to improved cancer detection performance.

\subsection{Error Analysis: Challenging Misclassification}

\begin{figure}[ht]
  \centering
  \includegraphics[width=\linewidth]{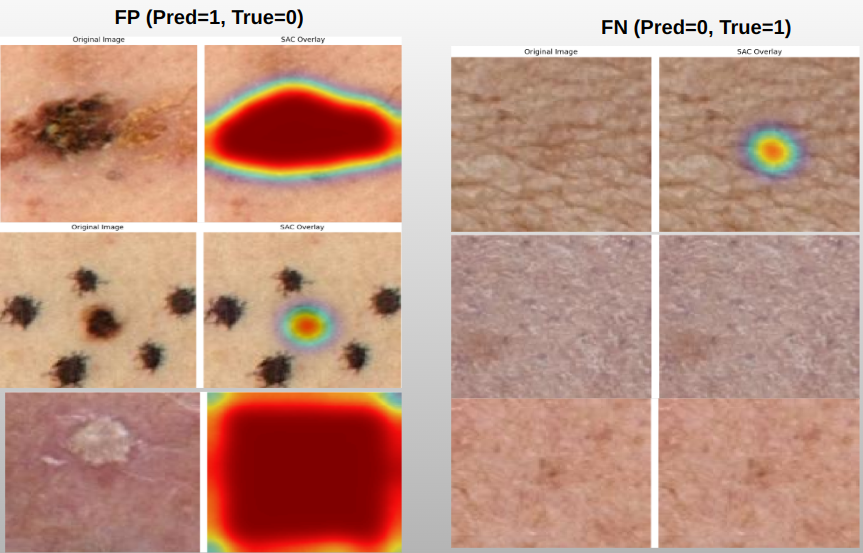}
  \caption{Failure case: despite localizing a lesion, the model overpredicts malignancy and misses a secondary lesion. This illustrates the limitations in handling multi-lesion frames and visually deceptive benign samples.}
  \label{fig:error_case}
\end{figure}

Figure~\ref{fig:error_case} showcases representative examples of false positive (FP) and false negative (FN) predictions by our segmentation-assisted classifier (EdgeNeXtSAC), along with their corresponding attention overlays. On the left panel, we examine three false positive cases where benign lesions were misclassified as malignant. In each instance, the attention heatmaps demonstrate that the model indeed focused on lesion regions. However, these lesions exhibit irregular borders, dark pigmentation, or structural complexity that resemble malignant patterns, misleading the classifier. In one case, the presence of multiple small nevi appears to have confused the model, which assigns high attention to a central lesion and mistakenly raises a malignancy score. Another case reveals the model's sensitivity to crusty textures that it interprets as malignant, reflecting the challenge of visually deceptive benign features.

On the right panel, we observe three false negative cases where malignant lesions were classified as benign. Notably, these lesions tend to be small, lightly pigmented, or lack high-contrast boundaries—properties that reduce their saliency to the model. In some cases, although the model correctly localizes the lesion (as seen from the SAC overlay), the low contrast and indistinct features result in an underestimated malignancy score. In one image, the lesion is barely perceptible to the human eye, further reinforcing the difficulty of detecting subtle malignancies, especially when such patterns are underrepresented in the training data.

These examples reveal two core limitations: (1) the classifier’s tendency to overfit to highly textured or dark benign patterns, and (2) its underperformance on subtle or low-contrast malignant lesions. They highlight the need for improved calibration, more diverse training examples (especially of hard negatives and faint malignancies), and potential incorporation of multi-lesion detection modules to handle spatial ambiguity and missed secondary lesions.

\subsection{Feature Importance}
An analysis of the average feature importance across all GBDT models reveals that several of the top-ranked predictors originate from our engineered feature set, validating the effectiveness of our feature engineering strategy. Notably, features such as \texttt{tbp\_lv\_deltaLBnorm}, \texttt{tbp\_lv\_eccentricity}, and \texttt{tbp\_lv\_symm\_2axis}—which capture lesion color asymmetry, geometric irregularity, and bilateral symmetry—appear among the top contributors to model performance. Traditional clinical attributes like lesion size (\texttt{clin\_size\_long\_diam\_mm}) and patient age (\texttt{age\_approx}) also retain high relevance. Importantly, the deep learning-derived prediction scores (\texttt{predictions\_eva} and \texttt{predictions\_edg}) are strongly ranked, reinforcing the benefit of integrating image-based softmax outputs into the tabular pipeline. In addition, lesion coordinates (\texttt{tbp\_lv\_x} and \texttt{tbp\_lv\_y}) and nevi-related confidence scores further contribute meaningfully, suggesting spatial and semantic lesion context are informative for malignancy prediction. Overall, the dominance of engineered and cross-modal features among the top contributors affirms the value of combining structured priors with learned representations in a unified ensemble framework.

\section{Conclusion}
\label{sec:conclusion}

This study presents a comprehensive and modular framework for skin lesion detection that addresses core challenges in dermatological AI, including class imbalance, poor generalization across diverse skin tones, and limited interpretability. Using the 3D-TBP dataset, our approach integrates segmentation-assisted classification, metadata-guided feature engineering, synthetic lesion augmentation via Stable Diffusion, and ensemble learning with CNN and GBDT models. The fusion of CNN softmax outputs with tabular features significantly boosts performance, achieving the highest partial AUC (0.1755), while segmentation-guided attention enhances spatial focus and explainability. Confidence distribution analysis revealed benign predictions tend to be confident, whereas malignant ones are more uncertain, reflecting data imbalance and lesion ambiguity. Visual inspection of misclassified samples highlighted difficulties in handling deceptive lesions and multi-lesion frames. Altogether, our system offers strong performance with interpretable outputs, fulfilling a growing clinical need for trustworthy AI in dermatology. Future work will expand this pipeline using transformer–CNN hybrids, lesion-grounded multimodal attention, improved class-conditional diffusion for rare class augmentation, and broader validation across ISIC and HAM10000 datasets.



{\small
\bibliographystyle{ieeenat_fullname}
\bibliography{main}
}

\end{document}